\documentclass[prl,showpacs,twocolumn]{revtex4}
\usepackage[dvips]{epsfig}

\newcommand{\cugeo}{CuGeO$_3$~}
\newcommand{\cumg}{Cu$_{1-x}$Mg$_x$GeO$_3$~}
\newcommand{\neel}{N\'{e}el~}
\newcommand{\tn}{\ensuremath{T_{\rm N}}~}

\newcommand{\figwidth}{0.85\columnwidth}

\begin{document}
 \title{Field-controlled phase separation at
the impurity-induced magnetic ordering \\
    in the spin-Peierls magnet \cugeo}
 \author{V. N. Glazkov}
 \email{glazkov@kapitza.ras.ru}
 \author{A. I. Smirnov}
 \affiliation{P. L. Kapitza Institute for Physical Problems RAS, 117334 Moscow,
 Russia}
\author{H.-A. Krug von Nidda}
\author{A. Loidl}
\affiliation{Experimental Physics V, Center for Electronic
Correlations and Magnetism, University of Augsburg,
    86135 Augsburg, Germany}
\author{K. Uchinokura}
 \altaffiliation[Present address: ]{The Institute of Physical and Chemical Research
     (RIKEN), Wako, Saitama 351-0198, Japan.}
\author{T. Masuda}
\altaffiliation[Present address: ]{Condensed matter sciences
division, Oak Ridge National Laboratory , Oak Ridge, TN 37831-6393,
USA.}

\affiliation{Department of Advanced Materials Science, The University
of Tokyo, 5-1-5 Kashiwa-no-ha, Kashiwa 277-8581, Japan}
 \date{\today}


 \begin{abstract}
The fraction of the paramagnetic phase surviving at the
impurity-induced antiferromagnetic order transition of the doped
spin-Peierls magnet Cu$_{1-x}$Mg$_x$GeO$_3$ ($x <$ 5\%) is found
to increase with an external magnetic field. This effect is
qualitatively explained by the competition of Zeeman energy and
exchange interaction between local antiferromagnetic clusters.
 \end{abstract}

 \maketitle

%\section{Introduction.}

Spin-chain magnets of spin-Peierls \cite{Hase} or Haldane
\cite{Haldane} types reveal a quantum-disordered singlet ground state
which is separated from the excited magnetic states by an energy gap.
Consequently, the disordered ground state is stable against the
influence of the crystal field or a weak interchain exchange.
Nevertheless, these quantum paramagnets may be driven into the
ordered state by means of weak doping, both magnetic as well as non
magnetic \cite{Hase3,Uchiyama}. Impurity-induced magnetic ordering is
explained by the local destruction of the singlet state and a
concomitant onset of local staggered magnetization. This local
antiferromagnetic (AFM) order consists of approximately 10 correlated
spins of a spin chain with a total magnetic moment of 1~$\mu_{\rm
B}$, see, e.g., Refs.~\onlinecite{Fukuyama,Miyashita}. According to
this model, the staggered magnetization attains a maximum value at
the sites of a spin chain near the impurity and decays exponentially
with the distance from the impurity. Such areas of the staggered
magnetization of about 10 interspin distances in size were detected
experimentally in a doped spin-Peierls magnet \cite{Kojima}.  The
overlap of the wings of these ordered areas and a weak interchain
exchange should result in long-range AFM order
\cite{Fukuyama,shender-kivelson,khomskii}.  Another theoretical model
\cite{Onishi} predicts a maximum of the AFM local order away from the
impurity, and NMR experiments \cite{Kikuchi} with off-chain doping of
the spin-Peierls compound CuGeO$_3$ revealed "the absence of
doping-induced moments next to an impurity". Apart from
one-dimensional magnets, impurity-induced ordering is also found in
the dimer spin-gap system TlCuCl$_3$
\cite{tlcucl3,tlcucl3-impurities-af}.

In spin-Peierls systems the spin-gap phase originates from the
magnetoelastic instability of a 3D crystal containing spin $S=1/2$
chains. At the temperature of instability, $T_{\rm SP}$, the chains
become dimerized with the exchange integral taking in turn two values
$J+\delta$ and $J-\delta$. The dimerization produces a gain in
exchange energy which exceeds the loss in elastic energy
\cite{Pytte}. The spin-Peierls transition and the impurity-induced
ordering were studied in detail in \cugeo having $T_{\rm SP} =
14.5$~K (see, e.g., Ref.~\onlinecite{glazkov-phasesep} and references
therein). The doping prevents dimerization, i.e. the temperature
$T_{\rm SP}$ of a doped crystal is lower, than that of a pure one,
and at an impurity concentration $x$ exceeding a threshold value
$x_{\rm c}$, the lattice remains uniform. Thus, depending on the
concentration, the N\'{e}el transition occurs on the dimerized (i.e.
spin-gap) or uniform (gapless) background. The ($T,x$)-diagram is
reported in a variety of papers, for the case of non magnetic
Mg-doping, with a detailed map of lattice and magnetic states given
in Ref.~\onlinecite{Mg&Zn-masuda}.

Because the staggered magnetization induced by an impurity is
localized on a short distance, the ordered phase should be highly
inhomogeneous. It was found experimentally
\cite{glazkov-phasesep,smirnov02,smirnov96} that below the N\'{e}el
temperature $T_N$ the AFM and paramagnetic (PM) resonance signals
coexist. Considering the spatial inhomogenity, it means that the
order parameter varies in space from a maximum value till zero  and a
true microscopic phase separation into AFM and PM phases occurs.
These experiments demonstrated the coexistence of the PM and AFM
responses in a magnetic field near $H = 12$~kOe. For nonzero
temperatures, the simple modeling (see
Ref.~\onlinecite{glazkov-phasesep}, Fig.~9) predicted the coexistence
of AFM areas of different sizes, surrounded by the residual of the
spin-gap matrix. According to this model, large AFM areas, ranging
for macroscopic distances, provide AFM resonance signals and an AFM
susceptibility. Small AFM areas provide PM resonance signals and a
Curie-like susceptibility due to their net magnetic moments. The
numerical simulation for a low-doped spin-Peierls system
\cite{Yasuda} confirmed a strongly inhomogeneous ground state with
practically 100\% spatial modulation of the absolute value of the
order parameter, this should imply a phase separation at a finite
temperature.

\begin{figure}
  \centering
  \epsfig{file=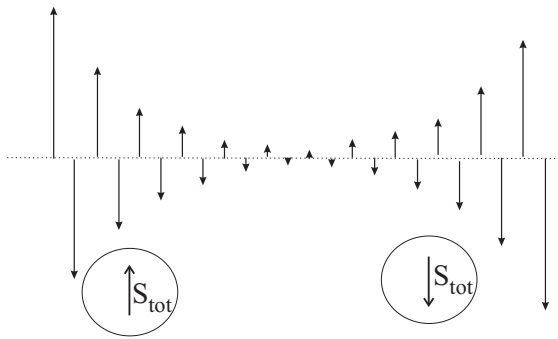, width=\figwidth, clip=}
  \caption{Schematic representation of the spin structure appearing in a spin-gap magnet on
  a spin-chain fragment with an odd number of spins. The arrows
  represent the average spin projections at the lattice sites. The local
  order parameters at the chain ends are correlated, but the net
  spins of the ordered areas are opposite.
  }\label{fig:clusters}
\end{figure}

In the present paper we investigate the influence of a magnetic field
on the structure of the impurity-induced ordered phase. The
motivation to perform these experiments is the hypothesis that the
correlation of the local order parameters of neighboring small AFM
areas should be broken by the magnetic field, if a parallel
orientation of the net magnetic moments hampers the coherent AFM
order in these two areas (see Fig.~\ref{fig:clusters}). Indeed, such
an influence of the magnetic field on the impurity-ordered phase can
be deduced from our magnetic resonance experiment, which
quantitatively probes the ratio of PM and AFM fractions dependent on
the field.

%\section{Samples and experimental details.}
\begin{figure}
  \centering
  \epsfig{file=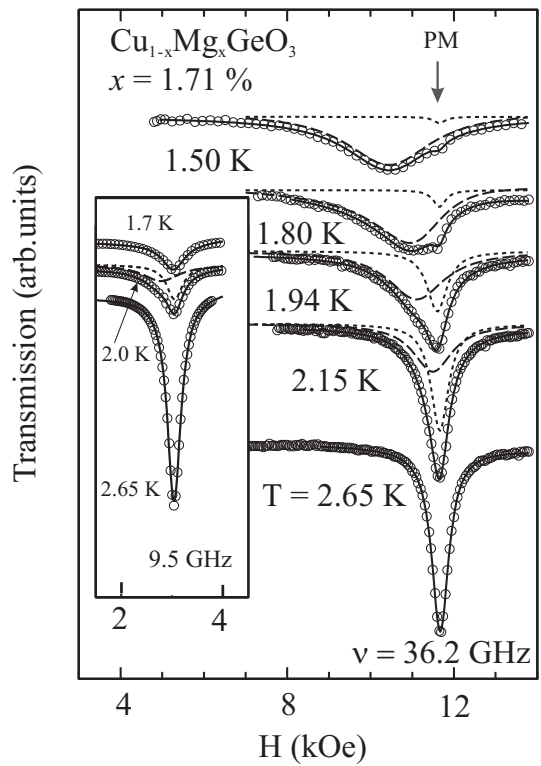, width=\figwidth, clip=}
  \caption{The temperature evolution of the ESR lineshape in a sample of \cumg
   with $T_{\rm N} = 2.25$~K,
  ${\bf H} \parallel b$.
  Position of the paramagnetic resonance is marked as "PM". Solid
  lines represent fitting by one or two Lorentzians, dashed and dotted lines are single-Lorentzian
  components slightly shifted for better clarity.
  }\label{fig:linesample}
\end{figure}

We used the same set of \cumg single crystals as for the
investigation of the phase diagram \cite{Mg&Zn-masuda} and of the
phase separation \cite{glazkov-phasesep}.
  Details of the sample
preparation and their quality control are described in
Ref.~\onlinecite{Mg&Zn-masuda}. The method of electron spin resonance
(ESR) enables one to detect the presence of two phases, because AFM
and paramagnetic phases have different resonance frequencies (or
different resonance fields, when measured at a fixed frequency $\nu$)
\cite{glazkov-phasesep}. The low-temperature ESR signal arises
predominantly from the impurities, because the susceptibility of the
spin-Peierls background is frozen due to the spin-gap. The
characteristic feature of the AFM-resonance mode is the nonzero
frequency in zero field. PM resonance of $S=1/2$ centers follows the
simple frequency-field relation $h \nu = g \mu_{\rm B} H$, and its
integral intensity is proportional to the static susceptibility,
hence, to the number of free spins. The fraction of the sample, which
remains paramagnetic below the N\'{e}el point, can be derived as the
ratio of the integral intensities of the PM resonance signals below
and above $T_{\rm N}$. We have performed the measurements of this
ratio in different magnetic fields by taking the ESR absorption as a
function of the magnetic field at different microwave frequencies. By
means of a set of transmission-type ESR spectrometers covering the
range from 18 to 40~GHz and using additionally a commercial Bruker
ELEXSYS X-Band spectrometer (9.5~GHz), we recorded the ESR lines in
the field range $3 \leq H \leq 12$~kOe.

%\section{Experimental results.}

As a typical example Fig.~\ref{fig:linesample} shows the evolution of
the ESR lineshape with temperature in Cu$_{1-x}$Mg$_x$GeO$_3$ ($x =$
1.71\%) obtained at a microwave frequency $\nu = 36.2$~GHz. Above the
\neel temperature $T_{\rm N} = 2.25$~K the absorption spectrum is
well described by a single Lorentzian line. An additional absorption
signal arises at lower fields, when the temperature decreases below
the \neel point. The resonance field of this additional signal
decreases with decreasing temperature. At $\nu > 18$~GHz it is
observable in the whole temperature range and its intensity grows at
cooling. Concomitantly, the intensity of the PM absorption decreases.
Thus, at low temperatures and at $\nu
> 18$~GHz the ESR absorption is dominated by the component with the
temperature-dependent resonance field. The frequency-field dependence
of this mode reveals the characteristics of an orthorhombic AFM and
can be identified as an AFM-resonance signal, see
Ref.~\onlinecite{glazkov-phasesep}. The \neel temperature $T_{\rm N}$
can be determined as the temperature of the onset of the
AFM-resonance absorption. The values of \tn obtained for all samples
in this way are in agreement with the results of magnetization
studies \cite{Mg&Zn-masuda}. The X-band (9.5~GHz) measurements (inset
of Fig.~\ref{fig:linesample}) reveal two coexisting lines only in a
narrow temperature interval below $T_N$, only the PM component is
visible at low temperatures. The AFM component disappears quickly
very close to $T_{\rm N}$ because the AFM-resonance gap becomes
larger than the microwave frequency. Nevertheless, the diminishing PM
component indicates the coexistence of the increasing AFM fraction.
Note that for a normal AFM transition only an AFM absorption signal
should be present below $T_{\rm N}$. For \cumg we observe this normal
scenario at $x>0.04$, where the dimerization is fully suppressed. At
low microwave frequencies, in contrast to the situation demonstrated
in the inset of Fig.~\ref{fig:linesample}, the ESR line in a normal
AFM disappears immediately below $T_{\rm N}$.

The concentration $x =$ 2.48\% lies in the crossover region, which
separates the N\'{e}el transition on the dimerized lattice from the
transition to the uniform AFM. For this concentration the
magnetization measurements show two cusps at temperatures $T_{\rm
N2}=3.85$~K and $T_{\rm N1}=3.15$~K, ascribed to the formation of the
AFM phase coexisting with short-range dimerization order and with
long-range spin-Peierls order, respectively  \cite{Mg&Zn-masuda}. We
found that the transformation of the ESR line, corresponding to AFM
ordering, occurs at the lower point $T_{\rm N1}$, only. For
temperatures $T
> T_{\rm N1}$ the ESR line consists of a single PM
resonance line.

\begin{figure}
  \centering
  \epsfig{file=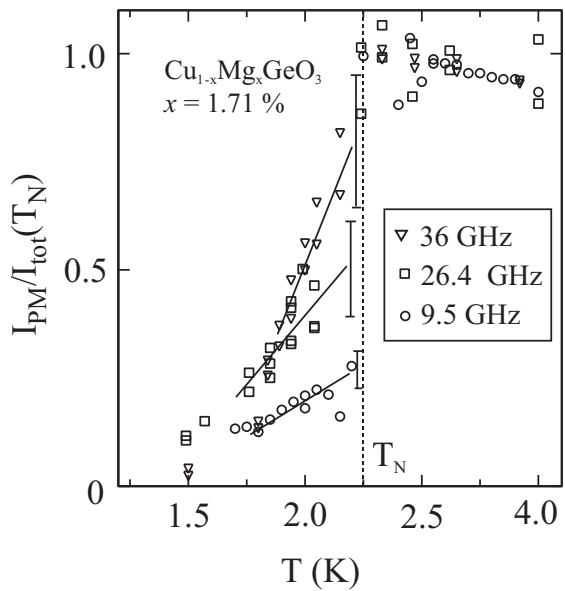, width=\figwidth, clip=}
  \caption{Temperature dependence of the
   integrated intensity of the paramagnetic ESR line. The vertical dotted line marks
  the \neel temperature. Solid lines indicate the extrapolations of the
  intensity of the paramagnetic component to the transition temperature.
  }\label{fig:i-below}
\end{figure}

\begin{figure}
  \centering
  \epsfig{file=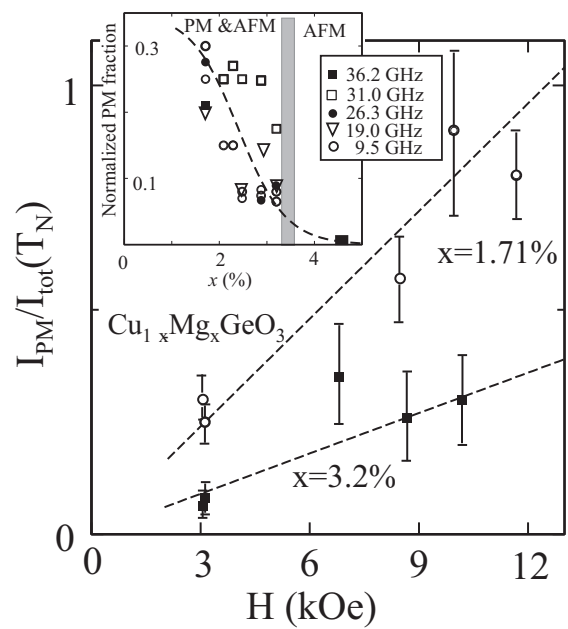, width=\figwidth, clip=}
  \caption{Field dependence of the PM fraction slightly below $T_N$.
  Dashed lines are guide to eye. Inset:
  the concentration dependence
  of the PM fraction.
 PM fractions, measured at different
 frequencies $\nu$ are renormalized by the
 multiplication factor 9.5~GHz$/\nu$.
 The dashed line is guide to the eye. The grey band marks the
 upper limit ($x=$3.5\%) of the concentration range, where the
 dimerization takes place \cite{Mg&Zn-masuda}.}\label{fig:i(h)}
\end{figure}

The ESR lines taken at different temperatures and frequencies were
numerically fitted by a sum of two Lorentzians, this fitting enables
one to measure the intensity of AFM (dashed line) and paramagnetic
resonances (dotted line in Fig.~\ref{fig:linesample}). The
temperature dependences of the integrated intensities of the
paramagnetic component are shown in Fig.~\ref{fig:i-below} for three
different microwave frequencies. Extrapolating the temperature
dependence of the intensity of the PM component to the transition
point allows the determination of the fraction of the sample
remaining paramagnetic at the \neel temperature. An unequivocal
determination of the two ESR components with close resonance fields
at temperatures just below $T_{\rm N}$ is rather uncertain and an
error of about 20\% of the total intensity occurs close to the
transition temperature for $\nu = 26.4$~GHz and $\nu = 36.0$~GHz. At
9.5~GHz, this error is only about 10\%. However, the observed
increase of the paramagnetic fraction with increasing microwave
frequency is significantly beyond this error, as shown in
Fig.~\ref{fig:i-below}. For  $x=$1.71\%  the PM fraction increases as
a function of frequency (i.e. field) from 0.2 at 9.5~GHz to 0.8 at
36~GHz. This field-dependence of the paramagnetic fraction at the
temperature just below $T_{\rm N}$ is given in Fig.~\ref{fig:i(h)}
for two concentration values. The relative intensity of the PM
component at the \neel point increases with increasing magnetic field
for both samples.

%\section{Discussion}

Concentration dependences of the PM fraction are shown in the inset
of Fig.~\ref{fig:i(h)}. The data taken at different frequencies are
normalized to the microwave frequency (i.e. to the ESR field in the
PM phase). The normalized dependences demonstrate a qualitatively
similar behavior with the PM fraction vanishing at $x \simeq 0.04$.
The concentration dependence of the PM fraction can be understood in
terms of the model developed previously in
Ref.~\onlinecite{glazkov-phasesep}: with increasing impurity
concentrations the volume of the PM phase should decrease, because
the distances between the impurities become shorter and the areas of
local AFM order more and more overlap by portions, where the spin
projections are large. Hence, the volume fraction remaining for the
undisturbed singlet matrix and for isolated AFM chain fragments
decreases. Besides that, at $x>$ 3.5\% the dimerization is suppressed
and neither a spin-gap nor phase separation shows up.

The magnetic field dependence of the PM fraction can be qualitatively
explained using the concepts of
Refs.~\onlinecite{Fukuyama,khomskii,shender-kivelson}. The local
order parameter decays exponentially with the distance from the
impurity, measured along the chain. At nonzero temperature the AFM
correlation of spin projections will be lost at a distance $L$ from
the impurity atom, this distance may be estimated by the relation
\begin{equation}\label{eqn:cluster-size-T}
 J S^2 \exp\left\{- \frac{L}{\xi} \right\} \sim k_{\rm B} T
\end{equation}
Here $J$ denotes the exchange-integral value and and $\xi$
characterizes the magnetic correlation length in the corresponding
direction. The ensemble of correlated spins in the chain fragment
of length $L$ near the impurity and of the spins from the
neighboring chains, correlated with the spins in the fragment due
to the interchain exchange, will be denoted further on as a
"cluster". As mentioned above, each cluster carries a net
magnetic moment of $1~\mu_{\rm B}$. As temperature decreases, the
size $L$ increases and the clusters begin to overlap, thus the
area of coherent AFM order increases. In a magnetic field the
Zeeman energy of the clusters attains a minimum, if the net
magnetic moments of all clusters are aligned parallel to the
field, but the exchange energy reaches a minimum at a coherent
correlation of local AFM-order parameters. Thus, an external
magnetic field destroys the AFM correlation of neighboring
clusters positioned as shown in Fig.~\ref{fig:clusters}. The
correlation of the order parameters of two clusters on a chain
fragment containing an odd number of spins will be destroyed by a
magnetic field given approximately by the relation
\begin{equation}\label{eqn:field-vs-coherence}
    g\mu_{\rm B}H \sim JS^2\exp\left\{-\frac{L}{\xi}\right\}
\end{equation}
Therefore, the magnetic field enlarges the number of clusters with
uncorrelated AFM order parameters and results in an increase of the
PM fraction of the sample. In other words, the increase of the field
should increase the PM fraction below the transition point, in
accordance with the experimental observations. This simplified
consideration does not include the interchain correlation which may
be also influenced by the magnetic field.

In this way, a strong magnetic field may destroy the long-range AFM
order, but the local order near the impurities will survive.
Therefore, this scenario can also explain the anomalously strong
field dependence of the \neel temperature reported recently for the
impurity-induced ordering in the Haldane magnet PbNi$_2$V$_2$O$_8$
\cite{masuda-tn(h)}.

%\section{Conclusions.}

In conclusion, we resume that the observation of a paramagnetic
ESR signal below the impurity-induced N\'{e}el transition reveals a
specific kind of AFM ordering, with a field-dependent microscopic
separation of magnetic phases. The fraction of the paramagnetic
phase is enlarged by application of an external magnetic field.
Within a qualitative model, considering the competition between
exchange interaction and Zeeman energy of the local
antiferromagnetic clusters formed around the impurities, the
influence of the magnetic field on the phase separation is
explained.

%\section{Acknowledgements.}

This work is supported by the Russian foundation for basic
research (RFBR) grant No. 03-02-16579, by the Deutsche
Forschungsgemeinschaft (DFG) via Sonderforschungsbereich SFB 484
(Augsburg) and via the joint project with RFBR under contract No.
436RUS113/628, and by the German Bundesministerium f\"ur Bildung
und Forschung (BMBF) under contract No. VDI/EKM 13N6917.

    %\bibliographystyle{article}
    %\selectlanguage{russian}
    %\bibliography{disser-bib,add-on}

\end{document}